\def\TeV{\ifmmode {\mathrm{\ Te\kern -0.1em V}}\else
                   \textrm{Te\kern -0.1em V}\fi}%
\def\GeV{\ifmmode {\mathrm{\ Ge\kern -0.1em V}}\else
                   \textrm{Ge\kern -0.1em V}\fi}%
\def\MeV{\ifmmode {\mathrm{\ Me\kern -0.1em V}}\else
                   \textrm{Me\kern -0.1em V}\fi}%
\def\pt{\ensuremath{p_{\mathrm{T}}}} 
\def\ipb{\mbox{pb$^{-1}$}}

\documentclass[preprint,10pt,5p]{elsarticle}

\usepackage{multicol,hyperref}

\usepackage{amssymb}
\usepackage{lineno}

\begin{document}

\begin{frontmatter}

\journal{Physics Letters B}
 
\title{Search for Heavy Long-Lived Charged Particles with the ATLAS Detector in $pp$ Collisions at $\sqrt{s} = 7\TeV$}

\author{The ATLAS Collaboration}

\address{}

\begin{abstract}
A search for long-lived charged particles reaching the muon spectrometer is performed using a data sample of $37~\ipb$ from $pp$ collisions at $\sqrt{s} = 7\TeV$ collected by the ATLAS detector at the LHC in 2010. No excess is observed above the estimated background.
Stable $\tilde{\tau}$ sleptons are excluded at 95\%~CL up to a mass of $136\GeV$, in GMSB models with $N_5=3$, ${m_{\rm{messenger}}=250\TeV}$, $\mathrm{sign}(\mu)=1$ and $\rm{\tan}\beta=5$. Electroweak production of sleptons is excluded up to a mass of $110\GeV$.
Gluino {\it{R}}-hadrons in a generic interaction model are excluded up to masses of $530\GeV$ to $544\GeV$ depending on the fraction of {\it{R}}-hadrons produced as $\tilde{g}$-balls. 

\end{abstract}

\begin{keyword}

SUSY \sep ATLAS \sep Long-Lived Particles

\end{keyword}

\end{frontmatter}

\section{Introduction}

Heavy long-lived particles (LLPs), with decay lengths longer than tens of meters, are predicted in a range of theories which extend the Standard Model. Supersymmetry (SUSY) models allow for meta-stable sleptons ($\tilde{l}$), squarks ($\tilde{q}$) and gauginos. Heavy LLPs produced at the Large Hadron Collider (LHC) could travel with velocity significantly lower than the speed of light. These particles can be identified and their mass, $m$, determined from their velocity, $\beta$, and momentum, $p$, using the relation $m=p/\gamma\beta$. Two different searches are presented in this paper, both use time of flight to measure $\beta$, and are optimized for the somewhat different experimental signatures of sleptons and {\it{R}}-hadrons.

Long-lived sleptons would interact like heavy muons, releasing energy by ionization as they pass through the ATLAS detector. A search for long-lived sleptons identified in both the inner detector (ID) and in the muon spectrometer (MS) is therefore performed. The results are interpreted in the framework of gauge-mediated SUSY breaking (GMSB)~\cite{comNote_GMSB} with the light $\tilde{\tau}$ as the LLP. If the mass difference between the other light sleptons and the light $\tilde{\tau}$ is very small, they may also be long-lived, otherwise the other light sleptons decay to the $\tilde{\tau}$.
 
Coloured LLPs ($\tilde{q}$ and $\tilde{g}$) would hadronize forming {\it{R}}-hadrons, bound states composed of the LLP and light quarks or gluons. They may emerge as neutral states from the $pp$ collision and become charged by interactions with the detector material, arriving as charged particles in the MS. A dedicated search for {\it{R}}-hadrons is performed in which candidates are required to have MS signals while ID and calorimeter signals are used if available. The ability to find {\it{R}}-hadrons without requiring an ID track makes this analysis complementary to the previous ATLAS paper searching for {\it{R}}-hadrons~\cite{comNote_rhadLimit}, that was based on ID and calorimeter signals without any requirement on the MS. In particular, the MS-based search presented here is more sensitive to models with larger $\tilde{g}$-ball fractions. Although the $\tilde{g}$-ball fraction is expected to be small~\cite{comNote_gball} we scan the full range in our analysis. The results of this analysis are interpreted in the framework of split SUSY~\cite{comNote_splitSusy} with the $\tilde{g}$ as the LLP.

\section{Data and simulated samples}

The work presented in this paper is based on $37~\rm{pb^{-1}} $ of $pp$ collision data collected in 2010. The events were selected online by muon triggers. Monte Carlo $Z \rightarrow\mu\mu$ samples are used for resolution studies. Monte Carlo signal samples are used to study the expected signal behavior and to set limits. The GMSB samples were generated with the following model parameters: the number of  super-multiplets in the messenger sector, $N_5=3$, the messenger mass scale, $m_{\rm{messenger}}=250\TeV$, the sign of the Higgsino mass parameter, $\rm{sign}(\mu)=1$ and the two Higgs doublets vacuum expectation values ratio, $\rm{tan}\beta=5$. The SUSY particle mass scale values, $\Lambda$, vary from 30 to 50 \TeV ~and the corresponding light $\tilde{\tau}$ masses from 101.9 to 160.7 \GeV. The $C_{grav}$ parameter was set to 5000 to ensure that the NLSP does not decay in the detector. The mass spectra of the GMSB models were generated by the SPICE program~\cite{comNote_spice} and the events were generated using Herwig~\cite{comNote_herwig}. The {\it{R}}-hadron samples were generated with $\tilde{g}$ masses from 300 to 700 \GeV. As discussed in Ref.~\cite{comNote_rhadLimit} several scattering and hadronization models can be used to describe the $\tilde{g}$ {\it{R}}-hadron spectrum and interactions with the detector material. Three different scattering models, the first described in Ref.~\cite{comNote_rhadInteraction}, the second in Ref.~\cite{comNote_rhadInteraction3} and the third in Ref.~\cite{comNote_rhadInteraction4}, and three different $\tilde{g}$-ball fractions (0.1, 0.5 and 1.0) are studied in this analysis. The different scattering models produce different fractions of candidates that arrive at the MS as charged particles while the $\tilde{g}$-ball fraction affects the number of candidates interacting as charged particles in the ID. All Monte Carlo events passed the full ATLAS detector simulation~\cite{comNote_geant4,comNote_atlasSimulation} and were reconstructed with the same programs as the data. All signal Monte Carlo samples are normalized to the integrated luminosity of the data, using cross-sections calculated to next to leading order, using the PROSPINO program~\cite{comNote_prospino}.

\section{The ATLAS detector}

The ATLAS detector~\cite{detector_paper} is a multipurpose particle physics apparatus with a forward-backward symmetric \\ cylindrical geometry and near $4\pi$ coverage in solid angle \footnote{ATLAS uses a right-handed coordinate system with its origin at the nominal interaction point in the centre of the detector and the $z$-axis coinciding with the axis of the beam pipe. The $x$-axis points from the interaction point to the centre of the LHC ring, and the $y$-axis points upward. Cylindrical coordinates ($r$, $\phi$) are used in the transverse plane, $\phi$ being the azimuthal angle around the beam pipe. The pseudorapidity is defined in terms of the polar angle $\theta$ as $\eta =  - \ln \tan(\theta/2)$.}. The ID consists of a silicon pixel detector, a silicon microstrip detector, and a transition radiation tracker. The ID is surrounded by a thin superconducting solenoid providing a 2 T magnetic field, and by high-granularity liquid-argon sampling electromagnetic calorimeters. An iron scintillator tile calorimeter provides hadronic coverage in the central rapidity range. The end-cap and forward regions are instrumented with liquid-argon calorimetry for both electromagnetic and hadronic measurements. The MS surrounds the calorimeters and consists of three large superconducting air-core toroids each with eight coils, a system of precision tracking chambers, and detectors for triggering. 

ATLAS has a trigger system to reduce the data taking rate from 40 MHz to $\sim$200 Hz, designed to keep the events that are potentially the most interesting. The first-level trigger (level-1) selection is carried out by custom hardware and identifies detector regions and a bunch crossing for which a trigger element was found. The high-level trigger is performed by dedicated software, seeded by data acquired from the bunch crossing and regions found at level-1. The components of particular importance to this analysis are described in more detail below.

\subsection{The muon detectors}

The MS forms the outer part of the ATLAS detector and is designed to detect charged particles exiting the barrel and end-cap calorimeters and to measure their momenta in the pseudorapidity range $|\eta| < 2.7$. It is also designed to trigger on these particles in the region $|\eta| < 2.4$. The chambers in the barrel are arranged in three concentric cylindrical shells around the beam axis at radii of approximately 5 m, 7.5 m, and 10 m. In the two end-cap regions, muon chambers form large wheels, perpendicular to the $z$-axis and located at distances of $|z|=$ 7.4 m, 10.8 m, 14 m, and 21.5m from the interaction point.

The precision momentum measurement is performed by Monitored Drift Tube (MDT) chambers, using the $\eta$ coordinate. These chambers consist of three to eight layers of drift tubes and achieve an average resolution of 80 $\rm{\mu} m$ per tube. In the forward region ($2 < |\eta| < 2.7$), Cathode-Strip Chambers are used in the innermost wheel. A system of fast trigger chambers, consisting of Resistive Plate Chambers (RPC) in the barrel region ($|\eta|<1.05$), and Thin Gap Chambers in the end-cap ($1.05 < |\eta| < 2.4$), delivers track information within a few tens of nanoseconds after the passage of the particle. The trigger chambers measure both coordinates of the track, $\eta$ and $\phi$.

When a charged particle passes through an MDT tube the electrons released by ionisation drift toward the wire. The hit radius is obtained from the hit time, using a known relation between the drift distance and the drift time. A segment is reconstructed as a line which is tangential to the cylinders of constant drift distance in the different layers. The drift time is estimated by subtracting the muon time of flight, $t_0$, from the measured signal time. Slow particles arrive at the MDT later than muons, and if this longer time of flight is not taken into account, the drift distances are overestimated.

The RPC chambers have an intrinsic time resolution of $\sim$1~ns while the digitized signal is sampled with a 3.12 ns granularity, allowing a measurement of the time of flight. When a charged particle passes through an RPC chamber the hit time and position are measured in the $\eta$ and $\phi$ directions separately.

\subsection{The tile calorimeter}

The tile calorimeter is a sampling calorimeter covering the barrel part of the hadronic calorimetry in ATLAS. It is situated in the region $2.3 < r < 4.3$ m, covering $|\eta| < 1.7$, and uses iron as the passive material and plastic scintillators as active layers. Cells are grouped radially in three layers. The tile calorimeter provides a timing resolution of 1-2 ns per cell for energy deposits typical of minimum-ionising particles (MIPs). The time measurement is described in detail in Ref.~\cite{comNote_tileNote}. The time of flight and hence the velocity of a candidate can be deduced from time measurements in the tile calorimeter cells along its trajectory. In this analysis, only cells with a measured energy deposition greater than 500 MeV are considered. The resolution of time measurements improves with increasing deposited energy.

\section{Reconstruction of Long-Lived Charged Particle Candidates}

Penetrating LLPs leave signals similar to muons except for their timing, and therefore their reconstruction is based on muon reconstruction. However, a late-arriving particle may be lost in standard muon reconstruction if its signals are not associated in time with the collision bunch crossing. Late arrival of the particle also spoils segment fitting in the MDTs.

The slepton search uses a dedicated muon identification package~\cite{comNote_muGirl} which starts from ID tracks and looks for corresponding hits in the MS, identifying candidates even when the segment reconstruction is imperfect, and refits the ID and MS hits in a combined track. Trigger detector hits arriving late with respect to the collision bunch crossing are also used. The next part of the LLP reconstruction is to estimate the particle velocity from the RPC, tile calorimeter and MDT~\cite{comNote_scientificNote}. The track is refitted after $\beta$ has been determined, resulting in a better momentum resolution since it uses a set of hits which are corrected to take into account the late arrival of the LLP at the different sub-detectors.

The {\it{R}}-hadron search employs a reconstruction method that relies only on the MS. The reconstruction is seeded by a feature found by the muon trigger, without requiring a match with the ID. This branch of the reconstruction collects hits and makes segments starting from the position and momentum of the trigger candidate in the middle station of the MS and extrapolates the track to the other stations. Once all segments are reconstructed, $\beta$ is estimated. A candidate which is not found by the muon trigger, i.e. if it arrives late at the trigger chambers and its hits are not associated with the collision bunch crossing, is not reconstructed in the MS-standalone method, leading to a loss of efficiency at low $\beta$.

\subsection{$\beta$ estimation}
The value of $\beta$ for each candidate is estimated by minimizing the total $\chi^2$ between the available timing measurements from the sub-detectors, and the timing expected from the hypothesized $\beta$ value. Contributions to the $\chi^2$ are calculated as follows.

\textbf{MDT segments}: An MDT segment is reconstructed as a line tangent to the circles of constant drift distance in the different layers, after the radii are estimated from the drift time using a known relation $R(t_{\rm{drift}})$, where $t_{\rm{drift}}$ is estimated as $t_{\rm{measured}}-t_0$. Slow particles have a longer time of flight, and a better segment fit is obtained with the correct $t_0$. For each test $\beta$, new MDT segments are built from a set of hits in a road around the extrapolated track, using the $t_0$ corresponding to the arrival time of a particle traveling with velocity $\beta$, $R(t_{\rm{measured}}-t_0(\beta))$. The $\chi^2$ representing the difference between the inferred position of the particle in the set of tubes and the segment position in the tube is minimized. For a low $\beta$ particle, this method recovers hits on segments that could be lost when fitting the segment assuming $\beta=1$. This estimate of the MDT $\beta$ is used in the {\it{R}}-hadron search. 

\textbf{MDT hits on track}: For the slepton search, the time estimate from the MDT segment method can be improved by performing a full track fit to the ID and MS hits. The estimated particle trajectory through each tube is significantly more accurate after the full track fit than in the segment finding stage. The time of flight of the particle to each tube is obtained using the difference between the time of flight corresponding to the refitted track position in the tube, $t_{{\rm{R}}}$ and the time actually measured, $t_0=t_{\rm{measured}}-t_{{\rm{R}}}$. The $\chi^2$ between the measured time of flight and the time of flight corresponding to the arrival time of a particle traveling with the test $\beta$ is minimized. 

\textbf{RPC and tile calorimeter}: The position and time are independent measurements for each hit (or cell). The $\chi^2$ minimization is performed using the measured times of hits on the candidate track. 

The time of flight measurement quality is sensitive to the time resolution of the detector. In a perfectly calibrated detector, any energetic muon coming from a collision in the interaction point will pass the detector at $t_{0}=0$. The $t_0$ distributions in the different sub-detectors are measured and their means used to correct the calibration. The observed width of these distributions after correction is used as the error on the time measurement in the $\beta$ fit and to smear times in the simulated samples.

The $\beta$ distributions of candidates obtained in the combined minimization are shown in Figure~\ref{fig:betaEstimation}, after the $\beta$-quality selection described in section~\ref{candidate_selection}. For the slepton search the mean $\beta$ value is 0.997 and the resolution is $\sigma_{\beta}=0.048$. For the {\it{R}}-hadron search the mean value is $\beta=1.001$ and the resolution is $\sigma_{\beta}=0.051$.

\subsection{Signal resolution expected in data}

\begin{figure}
  \centering
  \includegraphics[width=0.7\columnwidth]{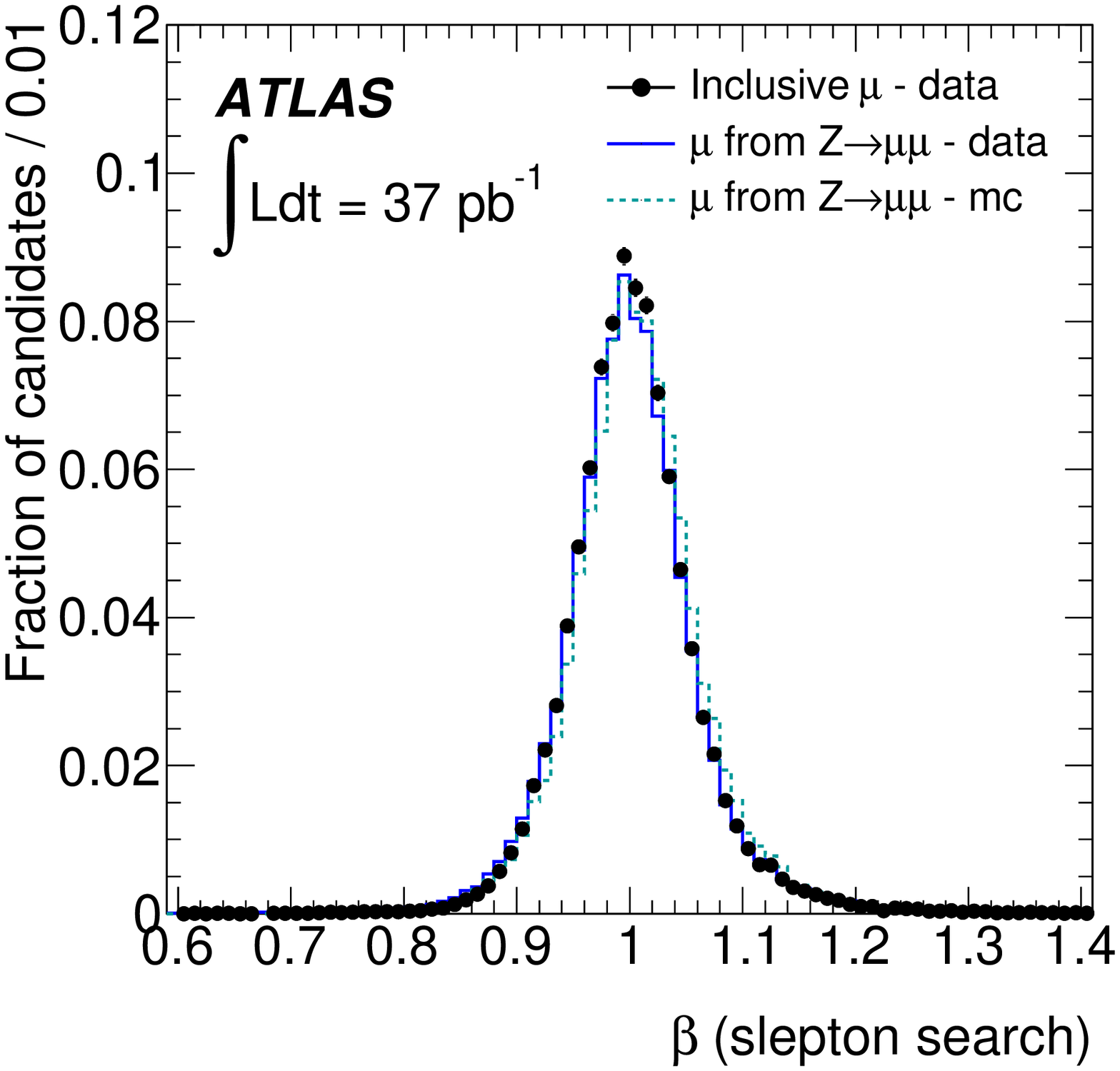}
  \includegraphics[width=0.7\columnwidth]{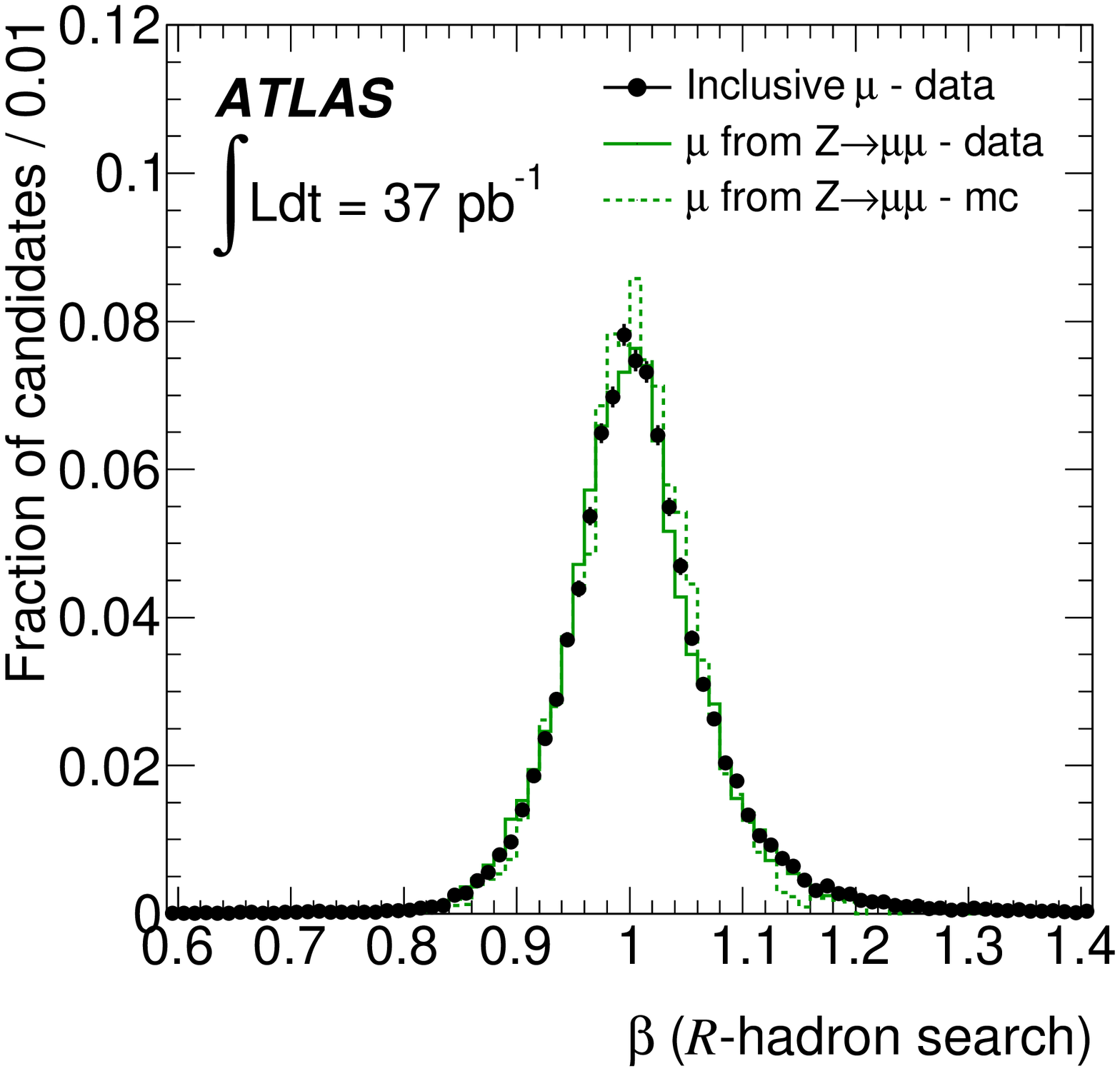}
  \caption{Distribution of $\beta$ for all candidates in data (points with error bars), muons from the decay $Z\rightarrow\mu\mu$ in data (full lines) and smeared Monte Carlo (dashed lines), in the estimation used in the slepton search (upper) and in the estimation used in the {\it{R}}-hadron search (lower). }
  \label{fig:betaEstimation}
\end{figure}

Since $\beta$ is estimated from the measured time of flight, for a given resolution on the time measurement, a slower particle has a better $\beta$ resolution. To simulate correctly the time resolution corresponding to the current state of calibration, hit times in simulated samples are smeared to reproduce the resolution measured in the data, prior to the $\beta$ estimation. Figure~\ref{fig:betaEstimation} shows the $\beta$ distribution for selected $Z\rightarrow\mu\mu$ decays in data and in Monte Carlo with smeared hit times. It can be seen that the smearing mechanism reproduces the measured muon $\beta$ distribution. The same time-smearing mechanism is applied to the signal Monte Carlo samples. 

\section{Candidate selection}
\label{candidate_selection}

\subsection{Trigger selection}

This analysis is based on events collected by two types of muon trigger chains. The trigger for the slepton search requires MS tracks to be matched with ID tracks in the high-level trigger. The estimated $p_{\rm{T}}$ is obtained from the combination of both systems, and is required to satisfy $\pt>13\GeV$. The trigger for the {\it{R}}-hadron search requires an MS-standalone muon trigger with $\pt>40$ GeV. The standalone triggers have less accurate \pt~estimates than the combined ones. 

The events are selected online by requiring at least one level-1 muon trigger. As level-1 muon triggers are accepted and passed to the high-level trigger only if assigned to the collision bunch crossing, late triggers due to late arrival of the particles are lost.  However, late triggers due to late arrival of slow particles from the collision bunch crossing are also lost. The level-1 trigger efficiency for particles arriving late at the MS is difficult to assess from data, where the overwhelming majority of candidates are muons. Therefore the trigger tracking efficiency is estimated from $Z \rightarrow \mu \mu$ data, while the effects of timing are estimated from simulated {\it{R}}-hadron and GMSB events passing the level-1 trigger simulation, which includes the timing requirements of the trigger electronics. The estimated trigger efficiencies for GMSB slepton candidates are between 80\% and 81\%. The {\it{R}}-hadron search is much more adversely affected by the loss of trigger efficiency for late candidates, since the reconstruction is seeded by the trigger, and so even if an event is triggered by another object, the candidate will be lost. 
For {\it{R}}-hadrons that could be reconstructed because they are charged in the MS, the trigger efficiencies range between 55\% for $m_{\tilde{g}} = 300\GeV$ to 38\% for $m_{\tilde{g}} = 700\GeV$. The estimated trigger efficiency with respect to all {\it{R}}-hadrons produced in the scattering model of Ref.~\cite{comNote_rhadInteraction} varies from 25\% for $m_{\tilde{g}} = 300\GeV$ to 17\% for $m_{\tilde{g}} = 700\GeV$. The effect of the trigger efficiencies can be seen in Tables~\ref{tab:bkg_rejection_cut_flow_CB} and~\ref{tab:bkg_rejection_cut_flow_MF} for the signal and data in the slepton and {\it{R}}-hadron searches respectively.

\subsection{Offline selection}
\label{secOfflineSelection}

Collision events are selected by requiring a good primary vertex with more than two ID tracks, and with $|z_{0}^{\mathrm{vtx}}|<150~\rm{mm}$ (where $z_{0}^{\mathrm{vtx}}$ is the $z$ coordinate of the reconstructed primary vertex).

Cosmic ray background is rejected by removing tracks that do not pass close to the primary vertex in $z$. For candidates with an associated ID track, candidates with $|z_{0}^{\mathrm{trk}} - z_{0}^{\mathrm{vtx}}|>10~\rm{mm}$ are removed, where $z_{0}^{\mathrm{trk}}$ is the $z$ coordinate at the distance of closest approach of the track to the origin. If no ID track is associated with the candidate, then it is still rejected if $|z_{0}^{\mathrm{trk}} - z_0^{\mathrm{vtx}}|>150~\rm{mm}$. Pairs of candidates with approximately opposite $\eta$ and $\phi$ are also removed.

The analysis searching for sleptons requires two candidates in each event, because two sleptons are produced, and both have a high probability to be observed in the MS. However, only one of them is required to pass the LLP selection. This requirement reduces background from $W$ production and QCD, but $Z \rightarrow \mu\mu$ decays remain. Any candidate that, when combined with a second muon, gives an invariant mass within 10~\GeV~of the $Z$ mass is rejected. In the {\it{R}}-hadron search, no requirement of two candidates per event is made, because {\it{R}}-hadrons may be neutral in the MS, or be lost by triggering in the next bunch crossing. Nevertheless, pairs consistent with the $Z$ mass are still rejected in the {\it{R}}-hadron search. The above requirements are grouped in Tables~\ref{tab:bkg_rejection_cut_flow_CB}  and~\ref{tab:bkg_rejection_cut_flow_MF} under the label ``event selection''.

The slepton search requires candidates to have $\pt > 40\GeV$, well above the efficiency plateau for the trigger threshold of $13\GeV$. A $p_{\rm{T}}$ requirement of $60\GeV$ is applied for all candidates in the {\it{R}}-hadron search, so as to be in the MS-standalone trigger-efficiency plateau. Candidates with $p_{\rm{T}}>1\TeV$ are rejected. This removes a few candidates with badly reconstructed momenta in both searches. Each candidate is required to have  $|\eta| < 2.5$. These requirements are grouped in Tables~\ref{tab:bkg_rejection_cut_flow_CB}  and~\ref{tab:bkg_rejection_cut_flow_MF} under the label ``candidate quality''.

The estimated $\beta$ is required to be consistent for measurements in the same sub-detector, based on the RMS of $\beta$ calculated from each hit separately. The estimated $\beta$ is also required to be consistent between sub-detectors. A $\beta$ measurement in at least two subdetectors is required for $|\eta|<1.7$. These requirements are grouped in Tables~\ref{tab:bkg_rejection_cut_flow_CB}  and~\ref{tab:bkg_rejection_cut_flow_MF} under the label ``$\beta$ quality''. Finally, in order to reject most muons, the combined $\beta$ measurement is required to be in the range $\beta < 0.95$.

\begin{table*}
  \begin{center}
  {\small
    \begin{tabular}{l |r| r r r r}
      \hline \hline
      &  & $\Lambda$ [\TeV] = 30 & $35$ & $40$ & $50$ \\ 
            	 &    data     & $m_{\tilde{\tau}}$ [\GeV] = $101.9$   & $116.3$    & $131.0$  & $160.7$   \\
            \hline\hline
 	Before selection	& -		  &  146.4		& 61.7 	& 28.7	&  7.3	 \\
       Trigger selection                         	 & 959921	  & 119.1 		& 50.4	& 23.3  	&  6.5	 \\
       Event selection                             	& 57382	  &  107.0		& 45.6	&  21.4 	& 6.0	  \\
       Candidate quality        & 5134	  & 91.4		& 38.8	&  18.3 	& 5.2	 \\
       $\beta$ quality             & 3470	 &  70.4		& 29.5	&  14.0	& 3.9	\\     
       $\beta < 0.95$   & 582	  & 51.8		& 21.7	& 11.2	 &3.0	\\
        \hline \hline
    \end{tabular} }
  \caption{Candidates in data compared to the simulated GMSB signal passing the selection stages in the slepton search. The Monte Carlo signal prediction is normalized to the data luminosity using the next to leading order cross-section.}
  \label{tab:bkg_rejection_cut_flow_CB}
  \end{center}
\end{table*}

\begin{table*}
  \begin{center}
  {\small
    \begin{tabular}{l |r| r r r r r}
      \hline \hline
       	 		   &    data     &$m_{\tilde{g}}$ [\GeV]$=300$   &  $400$    &  $500$  &  $600$ & $700$   \\ 
      \hline\hline
       Before selection		& -		& 4542		& 761		& 177.7	& 46.4	& 14.2    \\
       Trigger selection                         & 168043 &  1146	 	& 174		& 37.6	& 9.1		& 2.4		\\
       Event selection                            & 150771 &  1140		& 173		& 37.4	& 9.0		& 2.4		\\
       Candidate quality       & 6334	& 504		& 75		& 15.7	& 3.8		& 1.0		\\
       $\beta$ quality             & 4998	& 443		& 66		& 13.9	& 3.3		& 0.8		\\
       $\beta < 0.95$   & 830	& 420		& 64		& 13.5	& 3.2		& 0.8		\\   
       \hline \hline
    \end{tabular} }
  \caption{Candidates in data compared to the simulated {\it{R}}-hadron signal passing the selection stages in the {\it{R}}-hadron search. The Monte Carlo signal prediction for the sample with the scattering model of Ref.~\cite{comNote_rhadInteraction} and a $\tilde{g}$-ball fraction of 0.1 is normalized to the data luminosity.}
  \label{tab:bkg_rejection_cut_flow_MF}
  \end{center}
\end{table*}

\section{Background estimation}

The background is mainly composed of high \pt ~muons with mis-measured $\beta$. The estimation of the background mass distribution is made directly from the data and relies on two premises: that the signal to background ratio before applying requirements on $\beta$ is small and that the probability density function (pdf) for the $\beta$ resolution for muons is independent of the source of the muon and its momentum. 

For each muon candidate passing the $\beta$ quality requirement, a random $\beta$ is drawn from the muon $\beta$ pdf. If this $\beta$ is inside the signal range, $\beta < 0.95$, a mass is calculated using the reconstructed momentum of the muon and the random $\beta$. The statistical error of the background estimation is reduced by repeating the procedure many times for each muon and dividing the resulting distribution by the number of repetitions. The mass histogram obtained this way represents the background estimation.

The $\beta$ distribution is different in different detector regions for three main reasons: different $\eta$ regions are covered by different technologies ($|\eta| < 1.05$ for RPC, $|\eta| < 1.7$ for tile calorimeter, $|\eta| < 2.5$ for MDT); the time of flight method is more precise when the distance between the interaction point and the detector element is larger; the measurement in some regions of the detector is less precise due to fewer detector layers and magnetic field inhomogeneities. 
The background estimation is performed in $\eta$ regions so that the $\beta$ resolution within each region is approximately the same. The muon $\beta$ pdf in each $\eta$ region is given by the histogram of the measured $\beta$ of all muons in the region.  The regions also differ in the muon momentum distribution, since for any \pt ~cut, $p$ is larger as $\eta$ increases. Therefore the combination of $p$ with random $\beta$ is done separately in each region and the resulting distributions are added together.

The efficiency of the requirements described in section~\ref{secOfflineSelection} to reject cosmic rays is estimated from data collected with a cosmic muon trigger in the empty bunches and periods without collisions, dropping the requirement of a good primary vertex. The number of remaining cosmic ray muons are estimated from the number of candidates rejected by these requirements in the collision sample and the rejection efficiency. This results in $1.3\pm0.2$ cosmic rays in the {\it{R}}-hadron search sample. The estimated cosmic ray contamination in the slepton search sample is $0.7\pm0.2$ candidates.

%
\section{Systematic uncertainties}
%

Several possible sources of systematic uncertainty have been evaluated. 

\subsection{Signal yields}

The total experimental systematic uncertainty in the signal yields is 6\% on average. The sources and their individual contributions are described below. 

An uncertainty of  3.4\% is assigned to the measurement of the integrated luminosity~\cite{comNote_lumi}. 

The systematic uncertainty associated with the trigger selection is estimated in Ref.~\cite{comNote_triggerEfficiencies} to be 0.73\% (0.35\%) and 0.74\% (0.42\%) in the barrel (endcap) for the two trigger chains used in the slepton search. An uncertainty of 5\% is estimated for the {\it{R}}-hadron search using similar methods. The trigger simulation models in detail the timing requirements of the trigger electronics and the efficiency loss due to particles arriving late at the MS. Differences between data and Monte Carlo trigger efficiency due to the difference in time resolutions between data and simulation were tested and are negligible.

The signal $\beta$ resolution expected in the data is estimated by smearing the hit times according to the spread observed in the time calibration. The systematic uncertainty due to the smearing process is estimated by scaling the smearing factor up and down, so as to bracket the distribution obtained in data. A 6\% (2\%) systematic uncertainty is associated with the smearing process for the GMSB models in the barrel (endcap). For the $\tilde{g}$ {\it{R}}-hadrons, the effect of the smearing is negligible. 

The systematic uncertainties due to track reconstruction efficiency and momentum resolution differences between ATLAS data and simulation are estimated to be 0.5\% for GMSB events and between 0.8\% and 1.3\% for {\it{R}}-hadrons in the different hadronization and interaction models. 

\subsection{Background estimate}

A total of 15\%  (20\%) uncertainty on the background is estimated for the slepton ({\it{R}}-hadron) search resulting from individual contributions discussed below.

A systematic variation of the $\beta$ distribution within each of the detector regions used in the background estimation may lead to a systematic error on the background estimation. To quantify the variability of the $\beta$ distribution within a region and its effect on the background estimate, each region is sub-divided into smaller regions and the variation of their $\beta$ distribution is used as a variability estimate. This leads to the dominant uncertainty in the background estimate.

The possibility that a $\beta$-distribution dependence on the candidate momentum or source would result in a systematic uncertainty on the background estimate was tested. The candidates in each $\eta$ region were divided by their momentum into two bins with similar counts. The independence of the $\beta$ pdf from the source of the muons is confirmed using $Z\rightarrow\mu\mu$ samples. Estimating the background using the pdf from the low or high momentum bins or from muons from $Z\rightarrow\mu\mu$ results in negligible systematic uncertainties.

Finally, the background estimation is based on a limited statistics sample, that of all candidates that passed the candidate quality requirements, before the cut on $\beta$. The tail of the background mass distribution has a significant contribution from a few high momentum events, and a statistical error arises from this. In order to calculate the sensitivity of the limits to the statistics of the momentum distribution, the candidate sample was divided randomly into two samples and the background estimate derived from each  sample separately. The resulting uncertainty in the slepton search ranges from 1.04 candidates for $m > 90\GeV$ to 0.14 candidates for $m > 140\GeV$, while the errors from sample statistics in the {\it{R}}-hadron search are negligible. 

\subsection{Theoretical cross-sections}

The PROSPINO program~\cite{comNote_prospino} is used to calculate the signal cross-sections at next to leading order and two sources of theoretical systematic uncertainties were considered: the  renormalization and factorization scales are changed upward and downward by a factor of two. This results in a systematic error of 7\% for GMSB cross-sections and 15\% for $\tilde{g}$ cross-sections~\cite{comNote_rhadLimit}. The parton density functions of CTEQ6.6~\cite{comNote_cteq} are used, and the uncertainty due to variations in the parton distribution functions is estimated to be 5\%.


%
%
\section{Results}

\begin{figure}
  \centering
  \includegraphics[width=0.7\columnwidth]{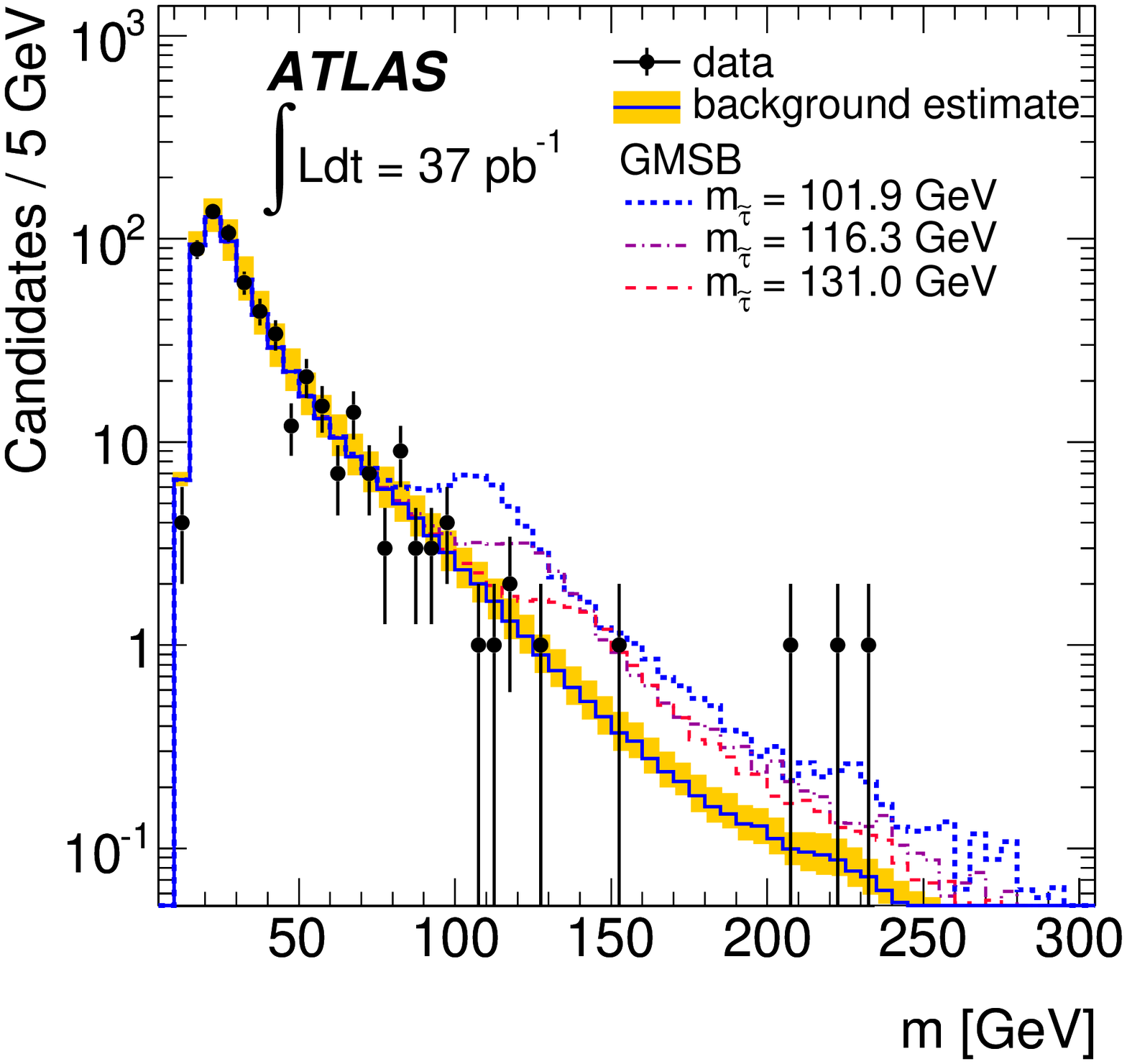}
  \includegraphics[width=0.7\columnwidth]{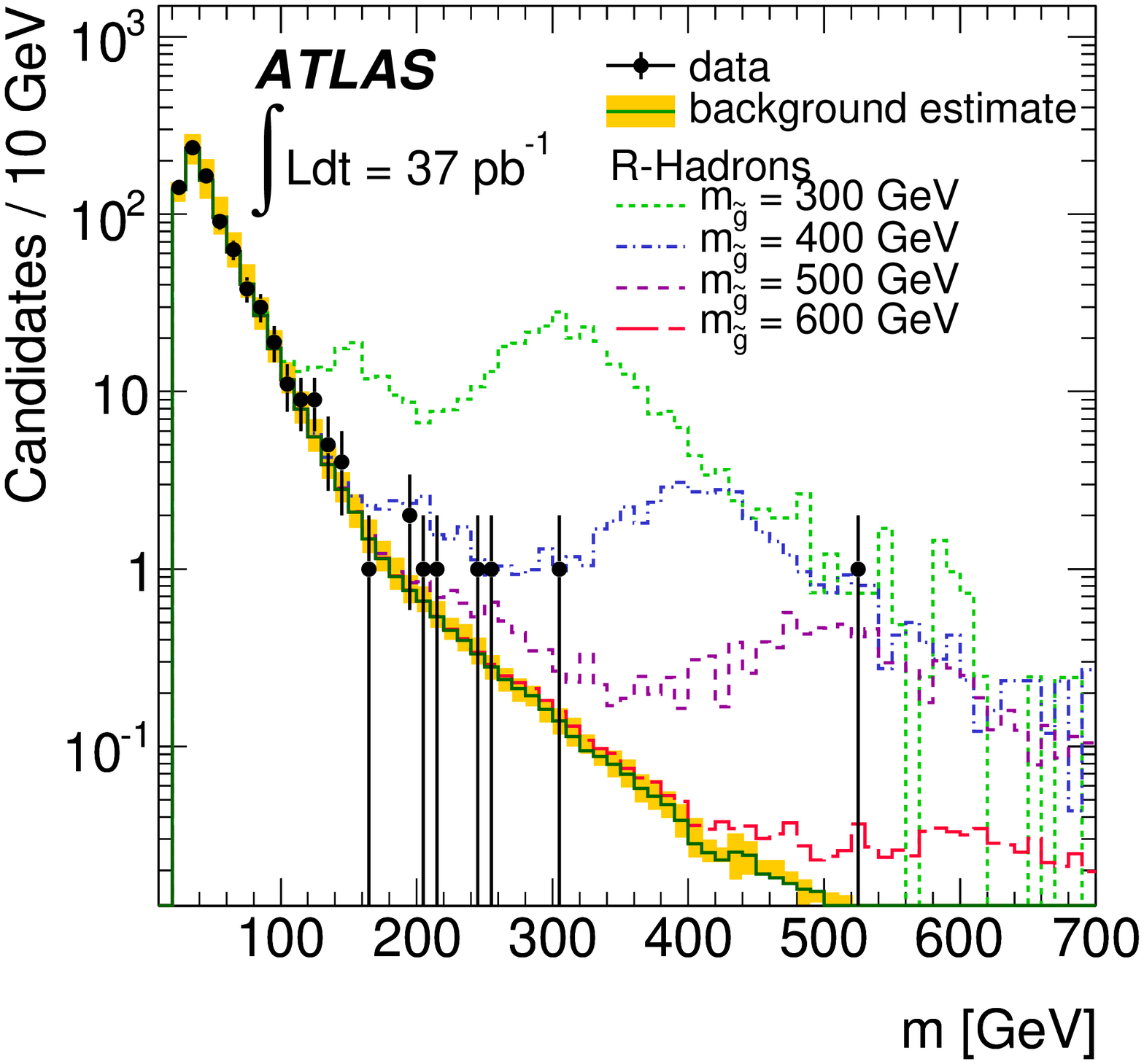}
  \caption{Candidate estimated mass distribution for data, expected background including systematic uncertainty, with simulated signals added, in the slepton (upper) and {\it{R}}-hadron (lower) searches.}
 \label{fig:signalOverEstimator}
\end{figure}

Figure~\ref{fig:signalOverEstimator} shows the candidate mass distribution for data and the estimated background with its systematic uncertainty. Good agreement is observed. The $\rm{CL_s}$ approach~\cite{comNote_CLs} for counting experiments is used to derive the limits for the production cross-section of GMSB slepton and $\tilde{g}$ {\it{R}}-hadron events. The limits are obtained by comparing the expected number of events with a candidate above a given mass cut with the actual number of events with a candidate above the same mass cut observed in the data. For each model, the mass cut is chosen to give the best expected limit. The mass cuts for the different models are summarized in Tables \ref{table:massCutGmsb} and \ref{table:massCutRhad} together with the expected signal and background in each case.

\begin{table}
  \centering
  {\small
\begin{tabular}{c c c c c}
\hline\hline
      & mass cut & expected& expected  &   \\
$m_{\tilde{\tau}} [\GeV]$         & [GeV]  &  signal &  background & data  \\
\hline\hline
101.9      & 90     & 35.9  & 19.2  & 16 \\
116.3      & 110    & 13.6  & 9.8  &  8  \\
131.0      & 120    & 7.3   &  7.2  &  5  \\
160.7      & 130    & 2.0   &  5.4  &  4  \\
\hline\hline
\end{tabular} }
  \caption{Mass cut and expected number of events as a function of the $\tilde{\tau}$ mass in the slepton search. The systematic uncertainties on the signal yield and background estimate are 6\%~and 15\%~respectively.}
  \label{table:massCutGmsb}
  \end{table}
  
\begin{table}
  \centering
  {\small
\begin{tabular}{c c c c c}
\hline\hline
      & mass cut & expected& expected  &   \\
      $m_{\tilde{g}} [\GeV]$          &  [GeV]  &  signal &  background & data  \\
\hline\hline
300  & 250  & 254.4 & 2.3 &  3	\\
400  & 350  & 36.2   &  0.7 & 1	\\
500  & 350   & 8.7    &  0.7 & 1	\\
600  & 350   & 2.2    &  0.7 &1	\\
700  & 350   & 0.6    &  0.7 & 1	\\
\hline\hline
\end{tabular} }
  \caption{Mass cut and expected number of events as a function of the $\tilde{g}$ mass in the {\it{R}}-hadron search. The systematic uncertainties on the signal yield and background estimate are 6\%~and 20\%~respectively.}
  \label{table:massCutRhad}
\end{table}

\begin{figure}
  \centering
  \includegraphics[width=0.7\columnwidth]{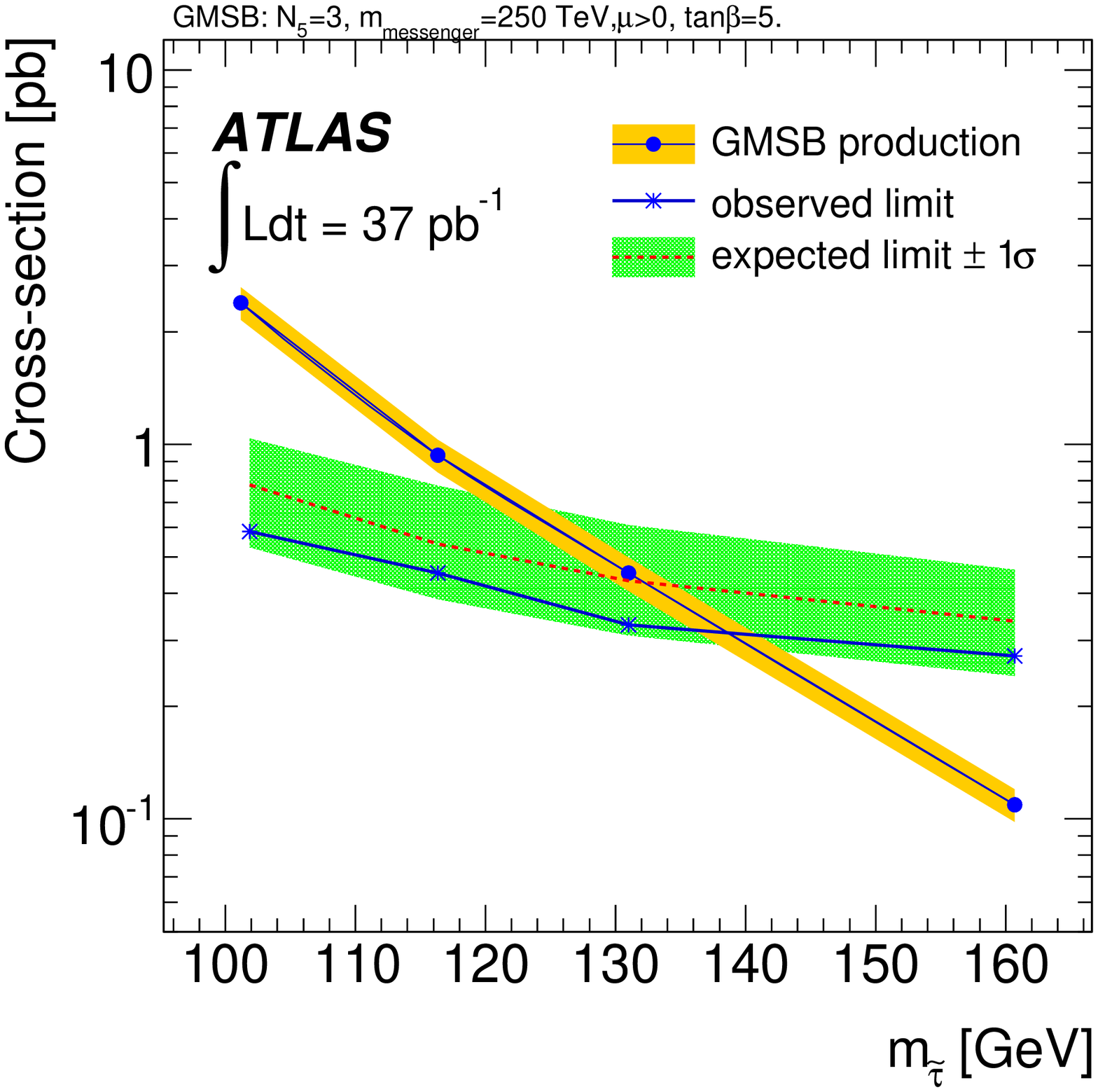}
  \includegraphics[width=0.7\columnwidth]{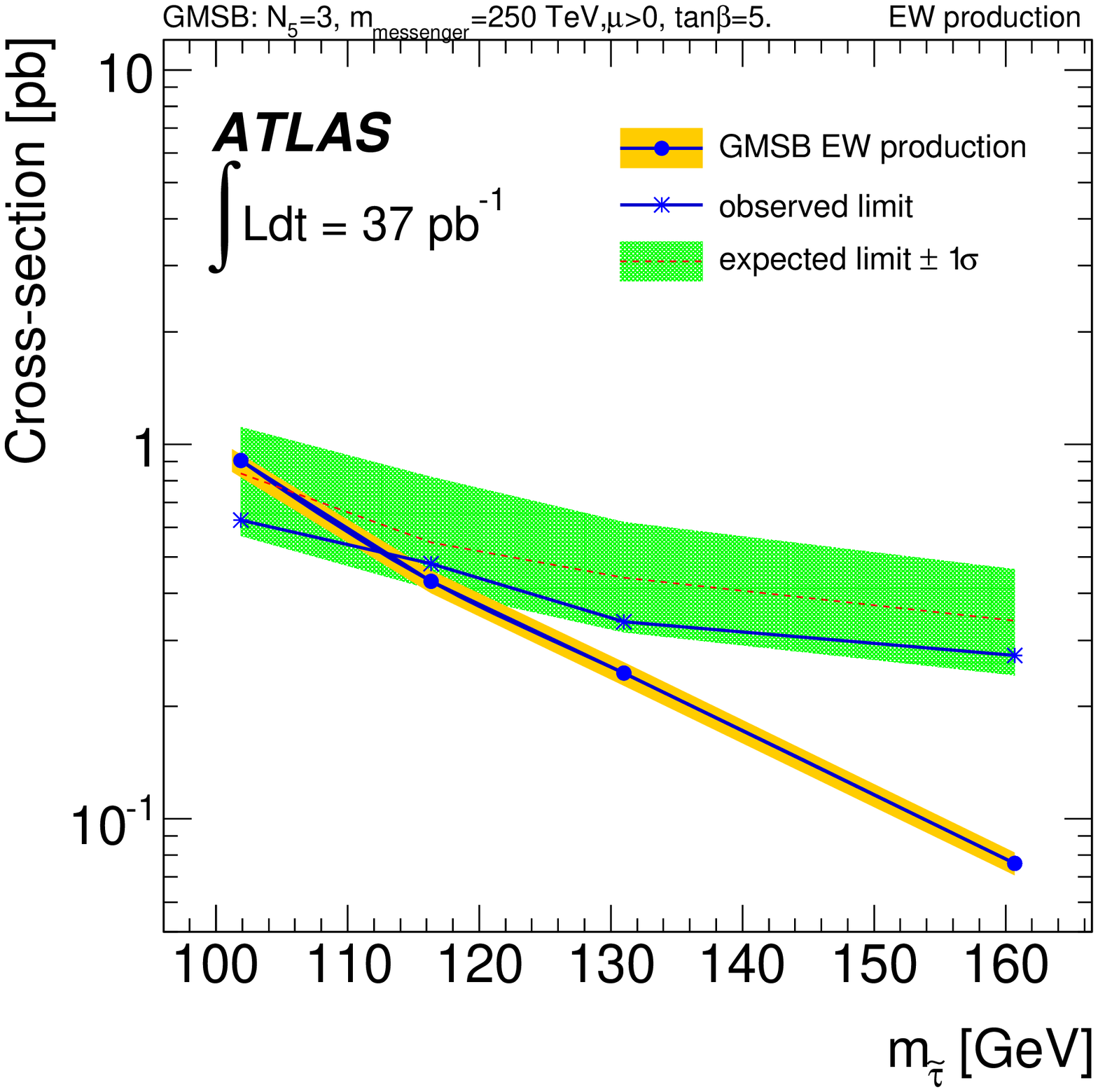}
  \caption{The expected production cross-section for GMSB events with $N_5=3$, $m_{\rm{messenger}}=250\TeV$, $\mathrm{sign}(\mu)=1$ and $\mathrm{tan}\beta=5$, and the cross-section upper limit at 95\%~CL  for the slepton search as a function of the $\tilde{\tau}$ mass (upper) and for sleptons produced in electroweak processes only (lower).}
 \label{fig:limits}
\end{figure}

The expected number of signal candidates for an integrated luminosity of $37~\ipb$ is added to the background estimation and compared to the data in Figure~\ref{fig:signalOverEstimator} for the GMSB and {\it{R}}-hadron models. The production cross-section for $\tilde{\tau}$ events and cross-section limit is shown in Figure~\ref{fig:limits}, as a function of the $\tilde{\tau}$ mass for the slepton search (upper). Stable $\tilde{\tau}$ are excluded at 95\%~CL up to a mass of $136\GeV$, in GMSB models with $N_5=3$, $m_{\rm{messenger}}=250\TeV$, $\mathrm{sign}(\mu)=1$ and $\mathrm{tan}\beta=5$. Figure~\ref{fig:limits} (lower) shows the limit obtained for sleptons produced only by electroweak processes, which have a smaller dependence on the model
parameters other than the slepton mass. Sleptons produced in electroweak processes are excluded up to a mass of $110\GeV$. Previous limits on stable sleptons are all below 100~\GeV~\cite{sleptonLimit}.  
These limits are only applicable to models where the $\tilde{\tau}$ or sleptons are the next to lightest SUSY particle, and their lifetime is sufficiently long to traverse the ATLAS experiment. In this case, the limits obtained for the above models are expected to have limited dependence on $\mathrm{tan}{\beta}$ and $N_5$.

\begin{figure}
  \centering
  \includegraphics[width=0.7\columnwidth]{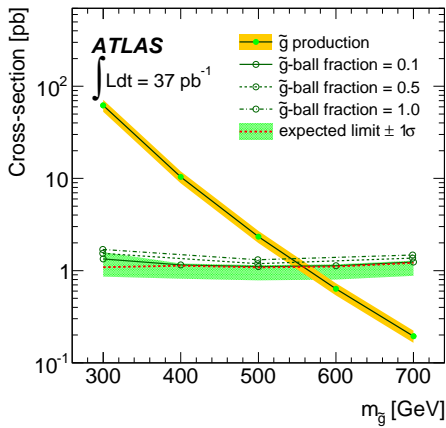}
  \caption{The expected production cross-section for {\it{R}}-hadron events and the cross-section limit at 95\%~CL as a function of the $\tilde{g}$ mass for the {\it{R}}-hadron search in the scattering model of Ref.~\cite{comNote_rhadInteraction} and for different $\tilde{g}$-ball fractions. The expected limit and its 1~$\sigma$ band are shown for 0.1 $\tilde{g}$-ball fraction.}
 \label{fig:RHLimits}
\end{figure}

Figure ~\ref{fig:RHLimits} shows the limits for $\tilde{g}$ {\it{R}}-hadrons in the scattering model of Ref.~\cite{comNote_rhadInteraction}. Such {\it{R}}-hadrons are excluded at 95\%~CL up to a mass of $544\GeV$ for a $\tilde{g}$-ball fraction of 0.1. Models with $\tilde{g}$-ball fractions of 0.5 and 1.0 are excluded up to masses of $537\GeV$ and $530\GeV$ respectively. Previous less stringent limits were set by the Tevatron~\cite{tevatronLimits} and by the CMS Collaboration~\cite{cmsLimit} which used the MS to select candidates, but not to measure $\beta$. For a $\tilde{g}$-ball fraction of 0.1, the ATLAS collaboration in Ref.~\cite{comNote_rhadLimit} sets limits that are higher then the limits presented here, not using the MS.  

%
%
\section{Summary and conclusion}

A search for long-lived charged particles reaching the muon spectrometer was performed with $37~\rm{pb^{-1}}$ of data collected with the ATLAS detector. No excess is observed above the estimated background and 95\%~CL limits on $\tilde{\tau}$ and {\it{R}}-hadron production are set.
Stable $\tilde{\tau}$'s are excluded up to a mass of $136\GeV$, in GMSB models with $N_5=3$, $m_{\rm{messenger}}=250\TeV$, $\mathrm{sign}(\mu)=1$ and $\mathrm{tan}\beta=5$. Sleptons produced in electroweak processes are excluded up to a mass of $110\GeV$.
Gluino {\it{R}}-hadrons in the scattering model of Ref.~\cite{comNote_rhadInteraction} are excluded up to masses of $530\GeV$ to $544\GeV$ depending on the fraction of {\it{R}}-hadrons produced as $\tilde{g}$-balls.

\section{Acknowledgements}

We thank CERN for the very successful operation of the LHC, as well as the
support staff from our institutions without whom ATLAS could not be
operated efficiently.

We acknowledge the support of ANPCyT, Argentina; YerPhI, Armenia; ARC, Australia; BMWF, Austria; ANAS, Azerbaijan; SSTC, Belarus; CNPq and FAPESP, Brazil; NSERC, NRC and CFI, Canada; CERN; CONICYT, Chile; CAS, MOST and NSFC, China; COLCIENCIAS, Colombia; MSMT CR, MPO CR and VSC CR, Czech Republic; DNRF, DNSRC and Lundbeck Foundation, Denmark; ARTEMIS, European Union; IN2P3-CNRS, \\ CEA-DSM/IRFU, France; GNAS, Georgia; BMBF, DFG, HGF, MPG
and AvH Foundation, Germany; GSRT, Greece; ISF, MINERVA, GIF, DIP and Benoziyo Center, Israel; INFN, Italy; MEXT and JSPS, Japan; CNRST, Morocco; FOM and NWO, Netherlands; RCN, Norway; MNiSW, Poland; GRICES and FCT,
Portugal; MERYS (MECTS), Romania; MES of Russia and ROSATOM, Russian
Federation; JINR; MSTD, Serbia; MSSR, Slovakia; ARRS and MVZT, Slovenia;
DST/NRF, South Africa; MICINN, Spain; SRC and Wallenberg Foundation,
Sweden; SER, SNSF and Cantons of Bern and Geneva, Switzerland; NSC, Taiwan;
TAEK, Turkey; STFC, the Royal Society and Leverhulme Trust, United Kingdom;
DOE and NSF, United States of America.

The crucial computing support from all WLCG partners is acknowledged
gratefully, in particular from CERN and the ATLAS Tier-1 facilities at
TRIUMF (Canada), NDGF (Denmark, Norway, Sweden), CC-IN2P3 (France),
KIT/GridKA (Germany), INFN-CNAF (Italy), NL-T1 \\ (Netherlands), PIC (Spain),
ASGC (Taiwan), RAL (UK) and BNL (USA) and in the Tier-2 facilities
worldwide.

\bibliographystyle{model1-num-names}
\bibliography{2010_11_paper}

\clearpage
\onecolumn

\begin{flushleft}
{\Large The ATLAS Collaboration}

\bigskip

G.~Aad$^{\rm 48}$,
B.~Abbott$^{\rm 111}$,
J.~Abdallah$^{\rm 11}$,
A.A.~Abdelalim$^{\rm 49}$,
A.~Abdesselam$^{\rm 118}$,
O.~Abdinov$^{\rm 10}$,
B.~Abi$^{\rm 112}$,
M.~Abolins$^{\rm 88}$,
H.~Abramowicz$^{\rm 153}$,
H.~Abreu$^{\rm 115}$
% ...
\bigskip

% ...
$^{10}$ Institute of Physics, Azerbaijan Academy of Sciences, Baku, Azerbaijan\\
$^{11}$ Institut de F\'isica d'Altes Energies and Universitat Aut\`onoma  de Barcelona and ICREA, Barcelona, Spain\\
% ...
$^{48}$ Fakult\"{a}t f\"{u}r Mathematik und Physik, Albert-Ludwigs-Universit\"{a}t, Freiburg i.Br., Germany\\
$^{49}$ Section de Physique, Universit\'e de Gen\`eve, Geneva, Switzerland\\
% ...
$^{88}$ Department of Physics and Astronomy, Michigan State University, East Lansing MI, United States of America\\
% ...
$^{111}$ Homer L. Dodge Department of Physics and Astronomy, University of Oklahoma, Norman OK, United States of America\\
$^{112}$ Department of Physics, Oklahoma State University, Stillwater OK, United States of America\\
% ...
$^{115}$ LAL, Univ. Paris-Sud and CNRS/IN2P3, Orsay, France\\
% ...
$^{118}$ Department of Physics, Oxford University, Oxford, United Kingdom\\
% ...
$^{153}$ Raymond and Beverly Sackler School of Physics and Astronomy, Tel Aviv University, Tel Aviv, Israel\\
% ...
\end{flushleft}

\end{document}